\documentclass[twocolumn]{aastex7}

\usepackage{mathtools}
\usepackage{amsmath}
\usepackage{graphicx}	

\begin{document}

\title{Super-Eddington Little Blue Dots May Reionize Helium Too Early}

\author[0000-0001-9420-7384]{Christopher Cain}
\email[show]{clcain3@asu.edu}
\affiliation{School of Earth and Space Exploration, Arizona State University, Tempe, AZ 85281, USA}

\author[0000-0002-0648-1699]{Brent Smith}
\email[]{bsmith18@asu.edu}
\affiliation{School of Earth and Space Exploration, Arizona State University, Tempe, AZ 85281, USA}

\author[0009-0007-0782-0721]{Gibson B. Bowling}
\email[]{gbbowlin@asu.edu}
\affiliation{School of Earth and Space Exploration, Arizona State University, Tempe, AZ 85281, USA}

\author[0000-0003-3307-7525]{Yongda Zhu}
\email[]{}
\affiliation{Steward Observatory, University of Arizona, 933 North Cherry Avenue, Tucson, AZ 85721, USA}

\author[0000-0003-1432-7744]{Lily Whitler}
\email[]{}
\affiliation{Kavli Institute for Cosmology, University of Cambridge, Madingley Road, Cambridge, CB3 0HA, UK}
\affiliation{Cavendish Laboratory, University of Cambridge, JJ Thomson Avenue, Cambridge, CB3 0US, UK}

\author[0000-0002-6150-833X]{Rafael Ortiz III}
\email[]{rortizii@asu.edu}
\affiliation{School of Earth and Space Exploration, Arizona State University, Tempe, AZ 85281, USA}

\author[]{Anson D'Aloisio}
\email[]{}
\affiliation{Department of Physics and Astronomy, University of California, Riverside, USA}

\author[0000-0001-8156-6281]{Rogier Windhorst}
\email[]{}
\affiliation{School of Earth and Space Exploration, Arizona State University, Tempe, AZ 85281, USA}

\begin{abstract}

The James Webb Space Telescope (JWST) has identified an abundant population of faint Active Galactic Nuclei (AGN) at $z \gtrsim 4$ which display broad emission lines, compact morphologies, and blue ultraviolet--optical continua. These ``Little Blue Dots' (LBDs) have been suggested to belong to the same family of objects as the more controversial ``Little Red Dots' (LRDs), with differences between the two largely owing to viewing angle effects. In this scenario, super-Eddington accretion resulting in high Extreme UV (EUV) and weak X-ray emission is invoked to explain the properties of both populations. We study the consequences of this super-Eddington scenario for the timing of Helium reionization. We find that observations which support an end to helium reionization no earlier than $z \approx 3$ disfavor scenarios in which the majority of observed $3 \lesssim z \lesssim 7$ LBDs are highly super-Eddington. For our fiducial super-Eddington accretion scenario, we find that the fraction of the LBD population in this state must be $\lesssim 10\%$, assuming LBDs make up $5\%$ of the $M_{ m UV} < -18$ galaxy population and have average escape fractions of $15\%$, comparable to recent observations. Our constraint assumes that LBDs dominate the Helium reionization budget, and would be tighter if bright quasars also contributed significantly. For a majority of LBDs to be super-Eddington, they would need to have small escape fractions ($\lesssim 1.5\%$) and/or be less abundant than observations suggest. Our conclusions are sensitive to the shape of the EUV spectra of super-Eddington black holes, motivating further study.

\end{abstract}


\section{Introduction}
\label{sec:intro}

In the last three years, JWST has identified a population of compact Broad-Line Active Galactic Nuclei (BLAGN) at $z \gtrsim 3$~\citep{Taylor2025,Juodzbalis2026,Zhang2026}.  
A large fraction of these objects have compact morphologies and blue UV-optical slopes, earning them the distinction of ``Little Blue Dots''~\citep[LBDs, ][]{Brazzini2026,Sneppen2026,Geris2026}, contrasting the better-known yet sub-dominant counterpart population, ``Little Red Dots''~\citep[LRDs, ][]{Matthee2024}, which share the blue UV slopes of LBDs but have red optical slopes resulting in a characteristic ``V-shaped'' SED.  
While LBDs are broadly recognized as un-obscured Type~I AGN, the physical nature of LRDs is much more contested.  
LRDs have many theories for the physical mechanisms leading to their unique SEDs and spectral signatures, including: obscuration of Type~I AGN~\citep{Greene2024,Madau&Maiolino2026}, ``black hole stars'' (BH*; i.e., accreting supermassive black holes (SMBHs) enveloped by a cocoon of thermalized gas in hydrostatic equilibrium,~\citealt{Naidu2025b, deGraaff2025}), or unique starbursts (i.e., globular clusters in formation with short-lived supermassive stars,~\citealt{Chisholm2026}).  

The focus of this work is a recent hypothesis, introduced by~\citep[][henceforth M26; see also~\citealt{Brazzini2026,Madau&Maiolino2026,Madau&Maiolino2026b}]{Madau2026} that LBDs and LRDs belong to the same parent population of BLAGN, the key difference between them being line-of-sight obscuration by the BH's torus.  
In this model, the peculiar features of these populations are explained by invoking super-Eddington accretion rates~\citep{Pacucci2024,Liu2025,Greene2026,Vaida2026,Begelman2026}.  
Specifically, some models for super-Eddington BH accretion predict strong emission in the extreme UV (EUV) together with weak X-ray emission.  This combination helps explain several key, shared features of LBDs and LRDs, including high-ionization emission lines~\citep{Zucchi2026}, the presence of broad H$\alpha$/H$\beta$ emission~\citep{Madau&Maiolino2026b,Scholtz2026}, steep UV slopes~\citep[$\beta_{\rm UV} \sim -2$;][]{Greene2024,Kocevski2025,Mascia2026}, and X-ray faintness~\citep{Yue2024,Ananna2024,Maiolino2025}.  

A consequence of such models is more efficient production of hydrogen (H) and helium (He)-ionizing photons than expected for standard galaxy and sub-Eddington AGN spectra.  
It was pointed out recently by~\citet{Su2026b} that one such class of models can produce hydrogen-ionizing photons up to $4\times$ more efficiently in faint AGN than standard power-law templates fit to the spectra of bright AGN/quasars~\citep{Lusso2015,DSilva2025a}. 
These results suggest that a similar conclusion may hold for He\,\textsc{ii}-ionizing photons with energies $> 4$ Ryd.  
If so, a super-Eddington explanation for LBDs and LRDs could have important implications for the double reionization of Helium at $z \sim 3$.  

Several lines of evidence indicate that He\,\textsc{ii} reionization must have ended not much earlier than $z = 3$.  
Measurements of the thermal history of the IGM indicate a distinct peak near this redshift, consistent with the expected heating from He\,\textsc{ii} ionizations~\citep{Becker2011,Boera2014,Villasenor2022}.  
Temperature measurements at higher redshifts are also lower than one would expect if He\,\textsc{ii}-reionization was concurrent with H \,\textsc{i} reionization~\citep{Daloisio2017, Walther2019,Boera2019,Gaikwad2020}. 
Observations of the He\,\textsc{ii} Ly$\alpha$ forest around $z \sim 3$ reveal large-scale opacity fluctuations consistent with an end to He\,\textsc{ii} reionization around this redshift~\citep{Worseck2011,Worseck2016,LaPlante2018}.  
Recently,~\citet{Gaikwad2025} argued that these fluctuations are consistent with a late and rapid end\footnote{Although some observations favor a gradual He\,\textsc{ii} reionization, with partial ionization extending up to $z \gtrsim 3.5$~\citep{Worseck2019,Makan2021}, observations generally do not support complete reionization of He\,\textsc{ii} earlier than $z \approx 3$.  } at $z < 3$, with a He\,\textsc{ii} fraction of tens of percent at $z \gtrsim 3.2$.  
This timing is consistent with the build-up of the quasar UV luminosity function (UVLF) around this redshift, suggesting bright quasars contributed significantly to this process~\citep{2016MNRAS.460.1885U, UptonSanderbeck2020}.  

In this {\it letter}, we study the implications of a super-Eddington LBD/LRD hypothesis for He\,\textsc{ii} reionization.   
In particular, we investigate whether constraints on He\,\textsc{ii} reionization permit the majority of LBDs to be super-Eddington at $z \gtrsim 3$, as required in such a scenario.  
This work is outlined as follows: \S\ref{sec:observations} briefly describes observational constraints on the UVLF of BLAGN and their ionizing escape fractions.  
In \S\ref{sec:AGNspectra}, we present our fiducial model for super-Eddington AGN spectra, and in \S\ref{sec:constraints} we compute the He\,\textsc{ii} reionization history and study the implications for LBD properties.  
We conclude in \S\ref{sec:conclusions}.  
Throughout, we assume the following cosmological parameters: $\Omega_m = 0.305$, $\Omega_{\Lambda} = 1 - \Omega_m$, $\Omega_b = 0.048$, $h = 0.68$, $n_s = 0.9667$ and $\sigma_8 = 0.82$, consistent with~\citet{Planck2018} results. Distances are in co-moving units unless otherwise specified. 

\section{Observations of Broad-Line AGN}
\label{sec:observations}

\begin{figure}
    \centering
    \includegraphics[width=\linewidth]{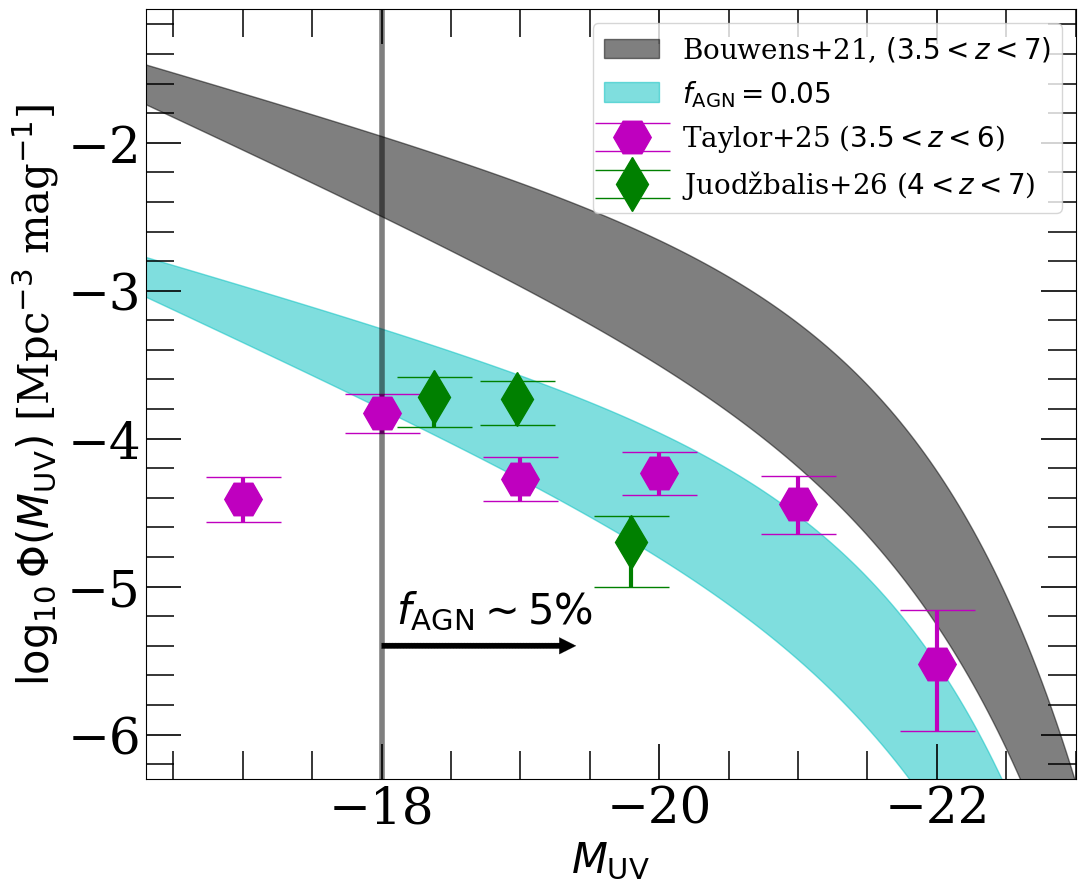}
    \caption{Measured UVLF of BLAGN at $3.5 < z < 7$ compared to that of the galaxy population at the same redshifts (gray).  The cyan-shaded region shows that an AGN fraction of $f_{\rm AGN} = 0.05$ is consistent with observations down to $M_{\rm UV} \sim -18$.  }
    \label{fig:UVLF}
\end{figure}

Recent JWST observations have constrained the fraction of UV-bright objects hosting BLAGN down to $M_{\rm UV} \approx -18$ at $3 \lesssim z \lesssim 7$~\citep{Harikane2023,Matthee2024,Grazian2024,Taylor2025,Juodzbalis2026}.  
In Figure~\ref{fig:UVLF}, we compare the measured BLAGN luminosity function at these redshifts to the galaxy UVLF.  
The green (magenta) points show recent measurements of the BLAGN UVLF from~\citet{Taylor2025, Juodzbalis2026}, and the black shaded region denotes the galaxy UVLF measured by~\citet{Bouwens2021} at $3.5 < z < 7$.  
The cyan-shaded region re-scales the latter by a factor of $0.05$, and overlaps the BLAGN measurements down to $M_{\rm UV} \approx -18$.  
These constraints motivate adopting a fiducial BLAGN fraction relative to all galaxies of  $f_{\rm AGN} = 0.05$ down to this magnitude cutoff\footnote{Observationally, one should subtract off LRDs from the BLAGN UVLF ($\approx10$--20\% at faint magnitudes, Figure 1 of~\citealt{Madau&Maiolino2026}).  However, in the super-Eddington scenario proposed in the same work, observed LRDs correspond to BLAGN viewed edge-on, suggesting that they would appear as LBDs from different orientations.  Furthermore, a greater-than-unity multiplicative correction may be needed to account for a lower-mass population of LRDs that would be observed as faint LBDs if viewed from a different angle (see Fig. 3 and surrounding discussion of~\citealt{Madau&Maiolino2026}.  )}.  

To model He\,\textsc{ii} reionization, we also need to assume an escape fraction for He\,\textsc{ii}-ionizing photons.    
While the escape fractions of the brightest AGN (quasars) are expected to be near unity, there is substantial uncertainty for the much fainter AGN considered here.  
Indeed, $f_{\rm esc}$ could be near unity or as low as percent-level depending on the geometry of the accretion disk and other physical conditions in the host galaxy; ~\citep[e.g.,][]{Trebitsch2018,Rosdahl2022}.  
\citet{Grazian2018} found that $z \sim 4$ AGN with $M_{\rm UV} \sim -23$ showed $f_{\rm esc} \sim 75\%$, while the observations of~\citet{Smith2024} at slightly lower redshifts ($2.5 < z < 4$) suggest a more modest $f_{\rm esc}$ for their Lyman-continuum (LyC) detections of $\lesssim 20\%$ (see also~\citealt{Smith2020}). 
Escape fractions close to $20\%$ were also reported in the recent work of~\citet{Mascia2026} for their bluest sources.  
In the model of M26, one might expect $f_{\rm esc}$ to be on the higher end of this range, since un-obscured viewing angles resulting in LBD-like properties account for the majority of the observed BLAGN population ($\sim 80$--$90\%$), and these lines of sight may have $f_{\rm esc}$ close to unity. 
On the other hand, it is possible that $f_{\rm esc}$ could fall off steeply as objects become dimmer, possibly reaching a few percent at the faint end (comparable to typical galaxy $f_{\rm esc}$; e.g.,~\citealt{Papovich2025}).   
Given these significant observational and theoretical uncertainties, we adopt an intermediate fiducial value of $f_{\rm esc} = 15\%$, and consider how deviations from this would affect our results. 

\section{Spectra of Super-Eddington AGN}
\label{sec:AGNspectra}

Our fiducial model for the spectra of super-Eddington BHs is that described by~\citet{Liu2002,Liu2003} (see also~\citealt{Liu2007}).  In their model, the geometry of the accretion flow can be described by a thin disk surrounded by a hot corona.   
\citet{Liu2003} proposed solutions in which the slim disk is gas or radiation-pressure dominated, which they referred to as the ``hard'' and ``soft'' states, respectively.  
The former is dominated by the hot corona and produces significant hard X-ray emission at $h_{\rm p}\nu \sim 1$--$10~$keV.  In the latter, there is little energy transferred to the corona, causing it to cool and largely shutting off hard X-ray emission, although the BH still emits significantly in the EUV.  
The observed X-ray faintness of LBDs is suggestive of something closer to the soft state solution~\citep[][see also~\citealt{Su2026a}]{Maiolino2025}.

\begin{figure*}
    \centering
    \includegraphics[width=\linewidth]{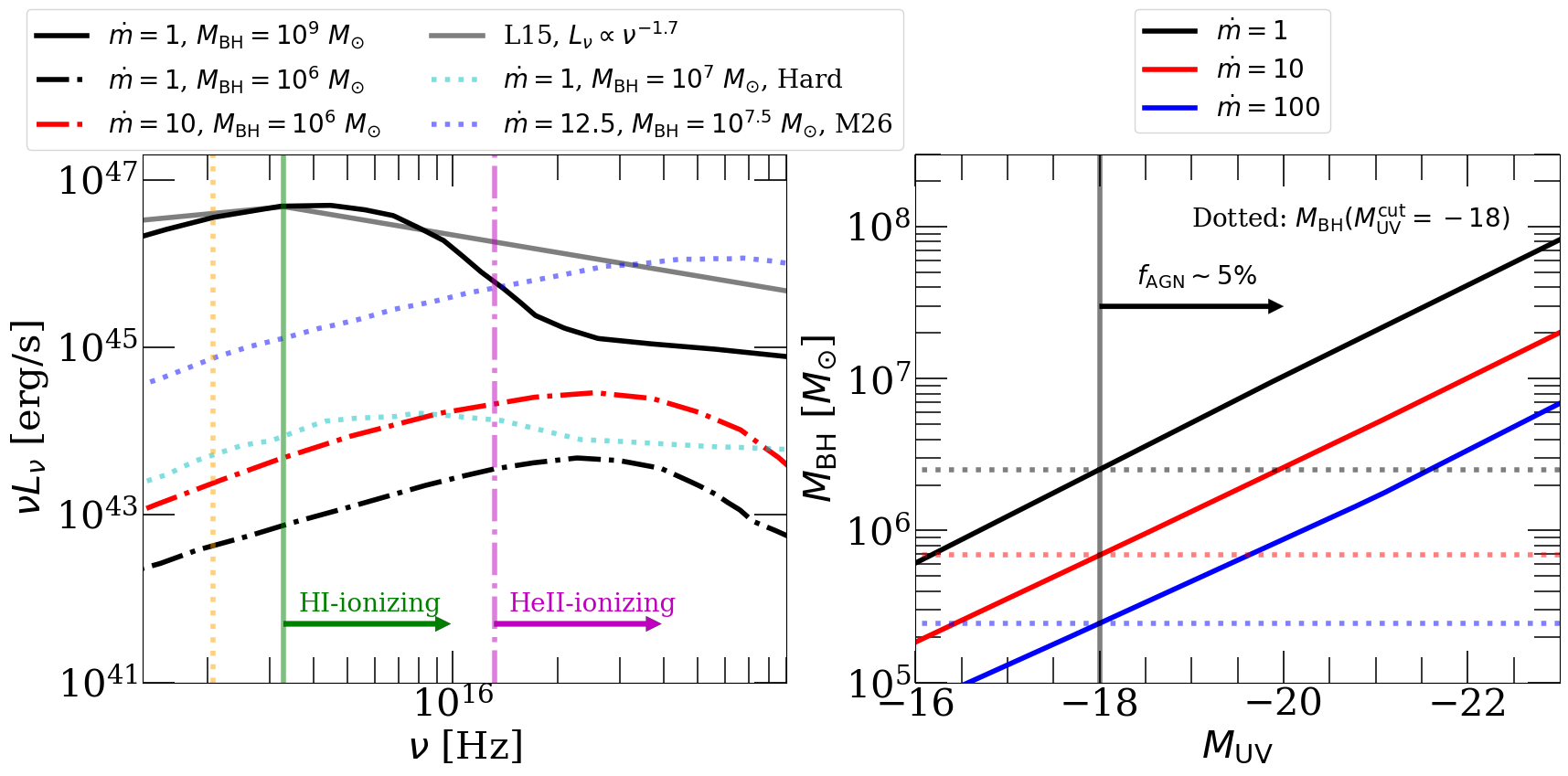}
    \caption{EUV SEDs for super-Eddington accretion models.  {\bf Left:} Shape of the ionizing SED for different BH masses and accretion rates.  The gray solid curve shows the L15 broken power-law template, and the black solid curve shows a soft state model for $M_{\rm BH} = 10^9~ \rm{M}_{\odot}$ and $\dot{m} = 1$.  Black and red-dashed curves show spectra for a $M_{\rm BH} = 10^6~\rm{M}_{\odot}$ BH with $\dot{m} = 1$ and $10$, respectively.  The vertical orange dotted line indicates $1450~\text{\AA}$, while the vertical green solid and magenta dot-dashed lines denote the ionizing energies of H\,\textsc{i} and He\,\textsc{ii}.  We also show, for reference, the SED of the hard state solution for $M_{\rm BH} = 10^7$ $M_{\odot}$ and $\dot{m} = 10$ (cyan dotted) and a model from M26 with $M_{\rm BH} = 10^{7.5}$ $M_{\odot}$ and $\dot{m} = 12.5$ (blue dotted).  {\bf Right:} $M_{\rm BH}$--$M_{\rm UV}$ relations for soft state solutions with $\dot{m} = 1$, $10$, and $100$.  The horizontal dotted lines denote where these intersect $M_{\rm UV} = -18$, to the right of which the observationally measured $f_{\rm AGN}$ is $\sim 5\%$.  These correspond to $M_{\rm BH} \sim 10^6~ \rm{M}_{\odot}$, for which production of H\,\textsc{i} and He\,\textsc{ii} ionizing photons is much more efficient than the L15 power law or BHs with much higher masses.  }
    \label{fig:spectra}
\end{figure*}

In both scenarios, the spectral energy distribution (SED) becomes harder in the EUV at lower BH masses.  
The left panel in Figure~\ref{fig:spectra} shows soft state SEDs for several values of the black hole mass $M_{\rm BH}$ and dimensionless accretion rate $\dot{m} \equiv \dot{M}/\dot{M}_{\rm Edd}$, where $\dot{M}_{\rm Edd}$ is the accretion rate at the Eddington limit. 
The gray solid line indicates the broken power-law SED inferred from AGN stacks in~\citet[][henceforth L15]{Lusso2015}.  
This SED has a spectral index $-0.6$ ($-1.7$) on the red (blue) side of the H\,\textsc{i} ionization edge, the latter being intermediate between the models in Figure 2 of~\citet{Madau2024}.  
The frequency corresponding to $1450~\text{\AA}$ is indicated by the orange dashed line, and the H\,\textsc{i} and He\,\textsc{ii}-ionizing thresholds are marked by solid green and dot-dashed magenta lines, respectively.  
The black solid curve shows a spectrum of a $\left( M_{\rm BH}, \dot{m} \right) = \left( 10^9~\rm{M}_{\odot},1 \right)$ BH, which matches well the L15 template to the left of the He\,\textsc{ii} ionizing edge and falls below it at higher energies.  
The black and red dot-dashed curves show spectra for a BH mass of $10^6~ \rm{M}_{\odot}$, assuming $\dot{m} = 1$ and $10$, respectively.  
The peak of the spectrum of these objects is shifted to much higher energies, above the He\,\textsc{ii} ionizing threshold\footnote{This scaling arises because of the $M_{\rm BH}^{-1/4}$ scaling of the accretion disk temperature in these models~\citep{Watarai2006}.  Some observational evidence for steeper scalings has been identified (e.g., by~\citealt{Windhorst2018}), which would result in even harder spectra at lower $M_{\rm BH}$ than we find here.  }.  
These smaller BHs thus produce He\,\textsc{ii} ionizing photons much more efficiently (per unit non-ionizing UV emission) than either the L15 power law or their more massive soft state counterparts.  
For reference, the dotted cyan and blue curves show the hard state spectrum from~\citet{Liu2003} with $\dot{m} = 10$ for a $10^7$ $M_{\odot}$ BH and the angle-averaged SED predicted for a $10^{7.5}$ $M_{\odot}$ BH with $\dot{m} = 12.5$ from M26.  
The shapes of these SEDs in the EUV are qualitatively similar to our fiducial soft state model.  

The right panel shows (for the soft state solution) what range of $M_{\rm BH}$ we should expect given observational constraints.  
The black, red, and blue solid lines show the $M_{\rm BH}$--$M_{\rm UV}$ relation predicted for the soft state solution assuming $\dot{m} = 1$, $10$, and $100$, respectively.  
Note that typical Eddington rates expected in the M26 model range from a few to tens, so we adopt $\dot{m} = 10$ as our fiducial value throughout.  
Unsurprisingly, at fixed $M_{\rm UV}$, $M_{\rm BH}$ decreases with increasing $\dot{m}$, since higher accretion rates produce higher luminosities per BH mass (although the relation is sub-linear for $\dot{m} \gg 1$).  
The vertical gray line denotes $M_{\rm UV} = -18$; at brighter magnitudes, observations are consistent with a $5\%$ AGN fraction (Figure~\ref{fig:UVLF}), and the dotted lines denote  where each relation crosses this threshold.  
The BH mass corresponding to $M_{\rm UV} = -18$ ranges from $\approx 2.5 \times 10^6 ~\rm{M}_{\odot}$ for $\dot{m} = 1$ to $\approx 2.5 \times 10^5 ~\rm{M}_{\odot}$ for $\dot{m} = 100$.  
Thus, {\it if the population of faint BLAGN is dominated by low-mass super-Eddington accretors, these will be much more efficient H\,\textsc{i} and He\,\textsc{ii} ionizing photon producers than expected given standard assumptions about AGN SEDs.}    

\section{Constraints from He\,\textsc{ii} Reionization}
\label{sec:constraints}

In this section, we model the He\,\textsc{ii} reionization history given different assumptions about the SEDs of high-$z$ BLAGN.  
The ionizing emissivity of the population is
\begin{equation}
    \label{eq:ndot_gamma}
    \dot{n}_{\gamma}^{\rm X} = f_{\rm esc}^{\rm X} f_{\rm Edd}^{\rm AGN} f_{\rm AGN} \int_{-\infty}^{M_{\rm UV}^{\rm cut}} \mathrm{d}M_{\rm UV} ~\xi_{\rm ion}^{\rm X}(M_{\rm UV},\dot{m})L_{\rm UV} \Phi(M_{\rm UV}) 
\end{equation}
where $X \in \{{\rm H\,\textsc{i}
, He\,\textsc{ii}
}\}$ is either neutral hydrogen or singly-ionized helium, $f_{\rm Edd}^{\rm AGN}$ is the fraction of AGN that are super-Eddington, $L_{\rm UV}$ is the non-ionizing UV luminosity, $\Phi(M_{\rm UV})$ is the galaxy UVLF, and $\xi_{\rm ion}^{\rm X}$ is the ionizing efficiency, given by
\begin{align}
    \label{eq:xi_ion_H}
    \xi_{\rm ion}^{\rm HI}(M_{\rm UV}[M_{\rm BH}],\dot{m}) = \frac{1}{L_{\rm UV}}\int_{\nu_0^{\rm HI}}^{{\nu_0^{\rm He\,\textsc{ii}}}} \mathrm{d}\nu ~ \frac{L_{\nu}(M_{\rm BH},\dot{m})}{h_{\rm p}\nu}
\end{align}
\begin{equation}
    \label{eq:xi_ion_He}
    \xi_{\rm ion}^{\rm He\,\textsc{ii}}(M_{\rm UV}[M_{\rm BH}],\dot{m}) = \frac{1}{L_{\rm UV}}\int_{\nu_0^{\rm He\,\textsc{ii}}}^{{20 \nu_0^{\rm He\,\textsc{ii}}}} \mathrm{d}\nu ~ \frac{L_{\nu}(M_{\rm BH},\dot{m})}{h_{\rm p}\nu}
\end{equation}
where $h_{\rm p}\nu_0^{\rm X}$ is the ionization energy of species $X$.  
We have truncated the second integral at $20\nu_0^{\rm He\,\textsc{ii}}$, or $\approx 1$ keV, which excludes hard X-rays.   
Hard X-rays are not produced in the soft state solution, which we prefer because it naturally explains the X-ray weakness of LBDs and LRDs.  
In the hard state case they would contribute an additional spatially uniform component to He\,\textsc{ii} reionization~\citep{UptonSanderbeck2020}, the inclusion of which would strengthen our conclusions (see below). 
The average escape fraction for photons that can ionize species X is given by Equation (13) of~\citet{Cain2025a}, 
\begin{equation}
    f_{\rm esc}^{\rm X} = \frac{\int_{-\infty}^{M_{\rm UV}^{\rm cut}} \mathrm{d}M_{\rm UV} ~ f_{\rm esc}^{\rm X}(M_{\rm UV}) ~\xi_{\rm ion}^{\rm X}(M_{\rm UV},\dot{m})L_{\rm UV} \Phi(M_{\rm UV}) }{\int_{-\infty}^{M_{\rm UV}^{\rm cut}} \mathrm{d}M_{\rm UV} ~\xi_{\rm ion}^{\rm X}(M_{\rm UV},\dot{m})L_{\rm UV} \Phi(M_{\rm UV}) }
\end{equation}
where $f_{\rm esc}^{\rm X}(M_{\rm UV})$ is the $M_{\rm UV}$-dependent escape fraction.  
For simplicity, we have assumed a constant $f_{\rm esc} = f_{\rm esc}^{\rm H\,\textsc{i}} = f_{\rm esc}^{\rm He\,\textsc{ii}}$.  
Since $\xi_{\rm ion}^{\rm He\,\textsc{ii}}$ in particular is a very strong function of $M_{\rm UV}$ for our preferred super-Eddington scenario, the average $f_{\rm esc}$ should be roughly the same as that close to $M_{\rm UV}^{\rm cut}$.  
We also note that the quantity $f_{\rm esc}^{X}(M_{\rm UV})$ is taken to be an isotropically-averaged quantity weighted by possible anisotropies in the intrinsic emission of the BH (see Figure 4 of~M26).  

It is possible for the ratio $f_{\rm esc}^{\rm He\,\textsc{ii}}/f_{\rm esc}^{\rm H\,\textsc{i}}$ to be higher or lower than unity, depending on the incident spectrum of ionizing radiation and the column density of absorbing material.  
Notably for our scenario,~\citet{Madau2024} pointed out that $f_{\rm esc}^{\rm He\,\textsc{ii}}$ can be a factor of $\approx 2$ lower than $f_{\rm esc}^{\rm H\,\textsc{i}}$ for gas with low H\,\textsc{i} column density through which most ionizing radiation is expected to escape, and this conclusion can hold even for spectra much harder than considered in that work (P. Madau, private communication).  
As such, we caution the reader that our $f_{\rm esc}$ parameter should be understood as the escape fraction for He\,\textsc{ii}-ionizing photons (which is most pertinent to this work), and that this may be lower than the better-studied H\,\textsc{i} escape fraction.  
In what follows, we will comment on how this distinction might affect our conclusions.

In the limiting case that faint AGN are the only source of He\,\textsc{ii} ionizing photons, the ionization history can be computed using the photon-counting formalism of~\citet{Madau1999}, 
\begin{widetext}
\begin{equation}
    \label{eq:photon_counting_eqn}
    \frac{\mathrm{d}x_{\rm He\,\textsc{iii}}}{\mathrm{d}t} = 
\begin{dcases} 
    \frac{\dot{n}_{\gamma}^{\rm He\,\textsc{ii}}}{n_{\rm He}} - C \alpha_{\rm B}^{\rm He\,\textsc{iii}}(T) n_{\rm e} x_{\rm He\,\textsc{iii}} & \text{if } x_{\rm He\,\textsc{iii}} < x_{\rm H\,\textsc{ii}} \\
    \frac{\dot{n}_{\gamma}^{\rm H\,\textsc{i}+He\,\textsc{ii}}}{n_{\rm H+He}} - C \alpha_{\rm B}^{\rm H\,\textsc{ii}+He\,\textsc{iii}}(T) n_e  x_{\rm He\,\textsc{iii}} & \text{if } x_{\rm He\,\textsc{iii}} = x_{\rm H\,\textsc{ii}}
    \end{dcases}
\end{equation}
\end{widetext}
where $x_{\rm He\,\textsc{iii}}$ is the mass-weighted He\,\textsc{iii} fraction,  $C$ is the clumping factor, $\alpha_{\rm B}^{\rm He\,\textsc{iii}}(T)$ is the case B recombination coefficient of He\,\textsc{iii} at temperature $T = 10^4$~K, $\alpha_{\rm B}^{\rm H\,\textsc{ii}+He\,\textsc{iii}}(T) \equiv [\alpha_{\rm B}^{\rm He\,\textsc{iii}}(T) n_{\rm He\,\textsc{iii}} + \alpha_{\rm B}^{\rm H\,\textsc{ii}}(T) n_{\rm H\,\textsc{ii}}]/[n_{\rm H\,\textsc{ii}} + n_{\rm He\,\textsc{iii}}]$ is the average recombination coefficient of H\,\textsc{ii} and He\,\textsc{iii}, and $\dot{n}_{\gamma}^{\rm H\,\textsc{i}+He\,\textsc{ii}} = \dot{n}_{\gamma}^{\rm H\,\textsc{i}} + \dot{n}_{\gamma}^{\rm He\,\textsc{ii}}$.  
We note that our temperature estimate is likely slightly low for the peak of He\,\textsc{ii} reionization~\citep{Villasenor2022}, and is thus a conservative choice since higher $T$ would lower the recombination rate.  
When He\,\textsc{ii} reionization lags behind H\,\textsc{i} reionization ($x_{\rm He\,\textsc{iii}} < x_{\rm H\,\textsc{ii}}$), it is a reasonable approximation that $> 4$ Ryd photons contribute exclusively to re-ionizing He\,\textsc{ii} and balancing He\,\textsc{iii} recombinations\footnote{The mean free path to He\,\textsc{ii} ionization is much higher at $4$ Ryd than that to H\,\textsc{i} ionization at redshifts of interest~\citep{Gaikwad2025}, suggesting that the majority of ionizations by $> 4$ Ryd photons in the ionized IGM should be of He\,\textsc{ii}.  }.  
If He\,\textsc{ii} reionization happens in lockstep with that of H\,\textsc{i} ($x_{\rm He\,\textsc{iii}} = x_{\rm H\,\textsc{ii}}$), excess $> 4$ Ryd photons will ionize H\,\textsc{i}, and the two energy ranges can be combined into a single $\dot{n}$.  
Since H\,\textsc{i} reionization is known to have ended by $z \approx 5.5$~\citep{Bosman2021,Gaikwad2023,Qin2024b}, we consider solutions in which $x_{\rm He\,\textsc{iii}} = x_{\rm H\,\textsc{ii}}$ only when $z_{\rm end}^{\rm He\,\textsc{iii}} > 5.5$;   
otherwise, we assume that $x_{\rm He\,\textsc{iii}} < x_{\rm H\,\textsc{ii}}$.   

\subsection{Effect on the He\,\textsc{ii} Reionization History}
\label{subsec:reion}

\begin{figure*}
    \centering
    \includegraphics[width=0.9\linewidth]{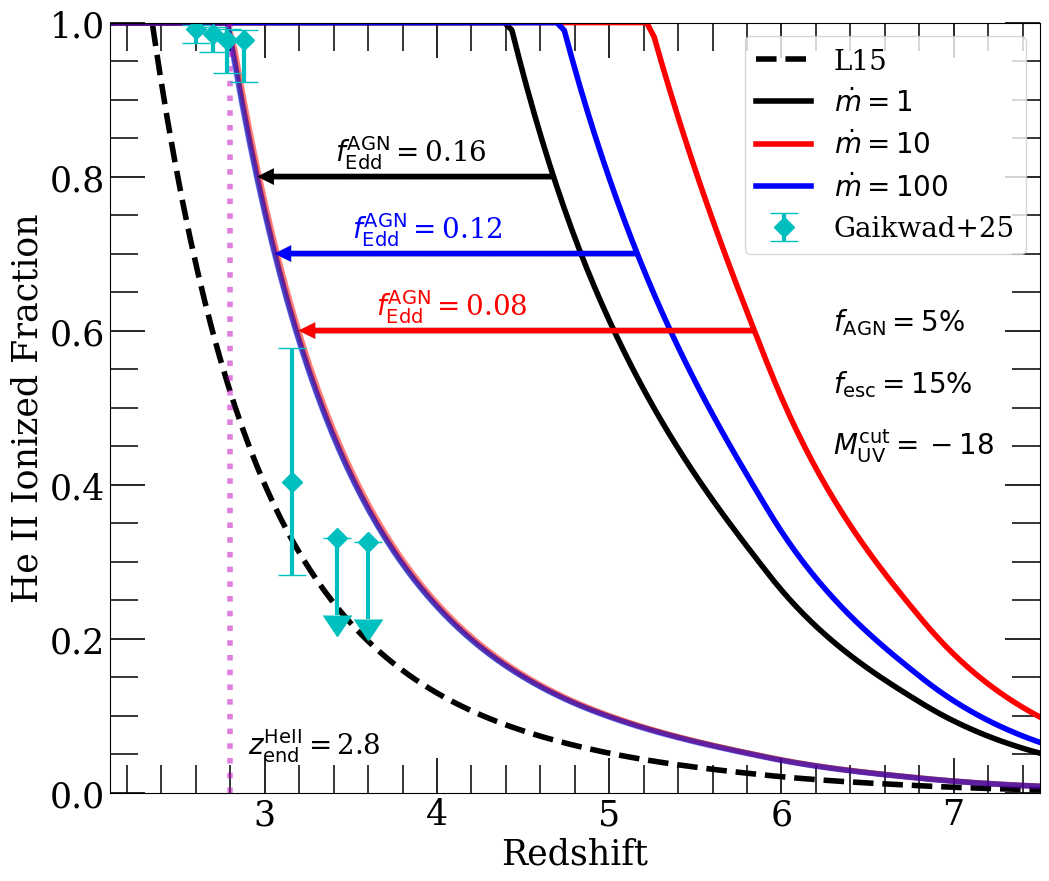}
    \caption{He\,\textsc{ii} reionization histories for different SED models assuming $f_{\rm AGN} = 0.05$, $f_{\rm esc} = 0.15$, and $M_{\rm UV}^{\rm cut} = -18$.  The dashed curve shows results for the L15 SED, while solid curves show scenarios for Soft State model with $\dot{m} = 1$, $10$, and $100$.  Respectively, completing He\,\textsc{ii} reionization at $z = 2.8$ (overlapping faded curves) requires $f_{\rm Edd}^{\rm AGN} = 0.16$, $0.08$, and $0.12$.  Cyan points denote constraints on the ionized fraction from~\citet{Gaikwad2025}.  }
    \label{fig:He II_reion_example}
\end{figure*}

In Figure~\ref{fig:He II_reion_example}, we show He\,\textsc{ii} reionization histories assuming the L15 SED (black dashed) and soft state super-Eddington models with $\dot{m} = 1$, $10$, and $100$ (solid curves), given $f_{\rm AGN} = 0.05$, $M_{\rm UV}^{\rm cut} = -18$, and $f_{\rm esc} = 0.15$ and a clumping factor $C = 3$.  
Cyan points denote recent measurements of the He\,\textsc{iii} fraction from~\citet{Gaikwad2025}, which suggest an endpoint of $z \approx 2.8$.  
We use this endpoint to place an upper limit on the allowed contribution to the He\,\textsc{ii} reionization budget from super-Eddington LBDs.  
We note that any scenario in which LBDs {\it under}-produce photons is considered permitted by the data, since the remaining photons could plausibly be supplied by other sources (such as bright quasars).  
The thick solid curves assume all AGN are super-Eddington with $f_{\rm Edd}^{\rm AGN} = 1$.  
The overlapping faded curves show the result of adjusting $f_{\rm Edd}^{\rm AGN}$ such that He\,\textsc{ii} reionization ends at $z = 2.8$ (vertical dotted line), in line with observations.  
For each curve we show the $f_{\rm Edd}^{\rm AGN}$ required to restore agreement with He\,\textsc{ii} reionization observations, which is between $\approx 0.08$--0.16 depending\footnote{The non-monotonic, factor-of-$2$ level shifts in the required $f_{\rm Edd}^{\rm AGN}$ owe to slight broadening of the super-Eddington SEDs with increasing $\dot{m}$ (see Figure 6 of~\citet{Su2026a}) together with shifts in the minimum $M_{\rm BH}$ required by observations (Figure~\ref{fig:spectra}).  } on $\dot{m}$ ($0.08$ for our fiducial value of $\dot{m} = 10$).  
Note that the L15 case ends He\,\textsc{ii} reionization at $z < 3$, suggesting that in this scenario BLAGN would not by themselves re-ionize He\,\textsc{ii} too early (consistent with the findings of~\citet{Madau2024}).   
Indeed, in this case, a higher $f_{\rm esc}$ and/or other sources of He\,\textsc{ii} ionizing photons would be required to re-ionize by $z = 2.8$.

We see that, given aforementioned assumptions, constraints on He\,\textsc{ii} reionization disfavor a scenario in which most BLAGN are super-Eddington at $z > 3$.  
This result is conservative for two reasons: (i) we have ignored the (expected) contribution to He\,\textsc{ii} reionization from bright quasars that are not counted in our galaxy UVLF~\citep{Finkelstein2022}, and (ii) we have ignored possible contributions from sub-Eddington faint AGN, which may be below current detection limits but could still contribute to the He\,\textsc{ii} ionizing budget~\citep{Su2026a}.  
Our results are consistent with the previous findings of~\citet[][their Figure 6]{Yoshiura2017a} that faint, low-mass, rapidly accreting BHs tend to re-ionize He\,\textsc{ii} much too early. 
Our work demonstrates that this would be the case if low-mass super-Eddington BHs dominate the BLAGN population observed by JWST at $3 < z < 7$, as suggested by M26 to explain LRDs.  
We note that our findings are specific to this proposed super-Eddington scenario, and do not necessarily apply to the more general scenario in which faint AGN drive He\,\textsc{ii} reionization, and possibly even H\,\textsc{i} reionization~\citep{Madau2015,Chardin2017,Daloisio2017,Garaldi2019,Madau2024}.  
Indeed, the L15 model itself is an example of a scenario in which faint AGN do not over-produce the He\,\textsc{ii} ionizing budget (see also~\citet{Madau2024}).  

While this work was in final preparation,~\citet{Mascia2026} reported constraints on the ionizing properties of compact broad-line emitters, including LRDs and LBDs.  
Using the \textsc{Sirocco} code~\citep{Matthews2025} to model intrinsic SEDs, they found modestly elevated H\,\textsc{i}-ionizing efficiencies $\log(\xi_{\rm ion}^{\rm H\,\textsc{i}}/[\text{Hz erg$^{-1}$}]) \sim 25.4$ and escape fractions $f_{\rm esc} \sim 0.2$ for their bluest sources.  
The former is somewhat lower than than the $\log(\xi_{\rm ion}^{\rm H\,\textsc{i}}/[\text{Hz erg$^{-1}$}]) \sim 26.5$ we find for an $M_{\rm UV} = -18$, Soft State $\dot{m} = 10$ LBD.  
However, it is not clear that their result is in tension with ours, since they do not report the $\dot{m}$ they infer for their objects.  
Indeed, we have checked that we recover  $\log(\xi_{\rm ion}^{\rm H\,\textsc{i}})$ similar to theirs for the Hard State solution with $\dot{m} = 0.1$, suggesting that the lower $\log(\xi_{\rm ion}^{\rm H\,\textsc{i}})$ they infer may indicate that their objects have $\dot{m} \ll 1$.
Their inferred escape fraction is consistent with our fiducial value.  

\subsection{Implications for AGN Properties}
\label{subsec:constraints}

\begin{figure*}
    \includegraphics[width=\linewidth]{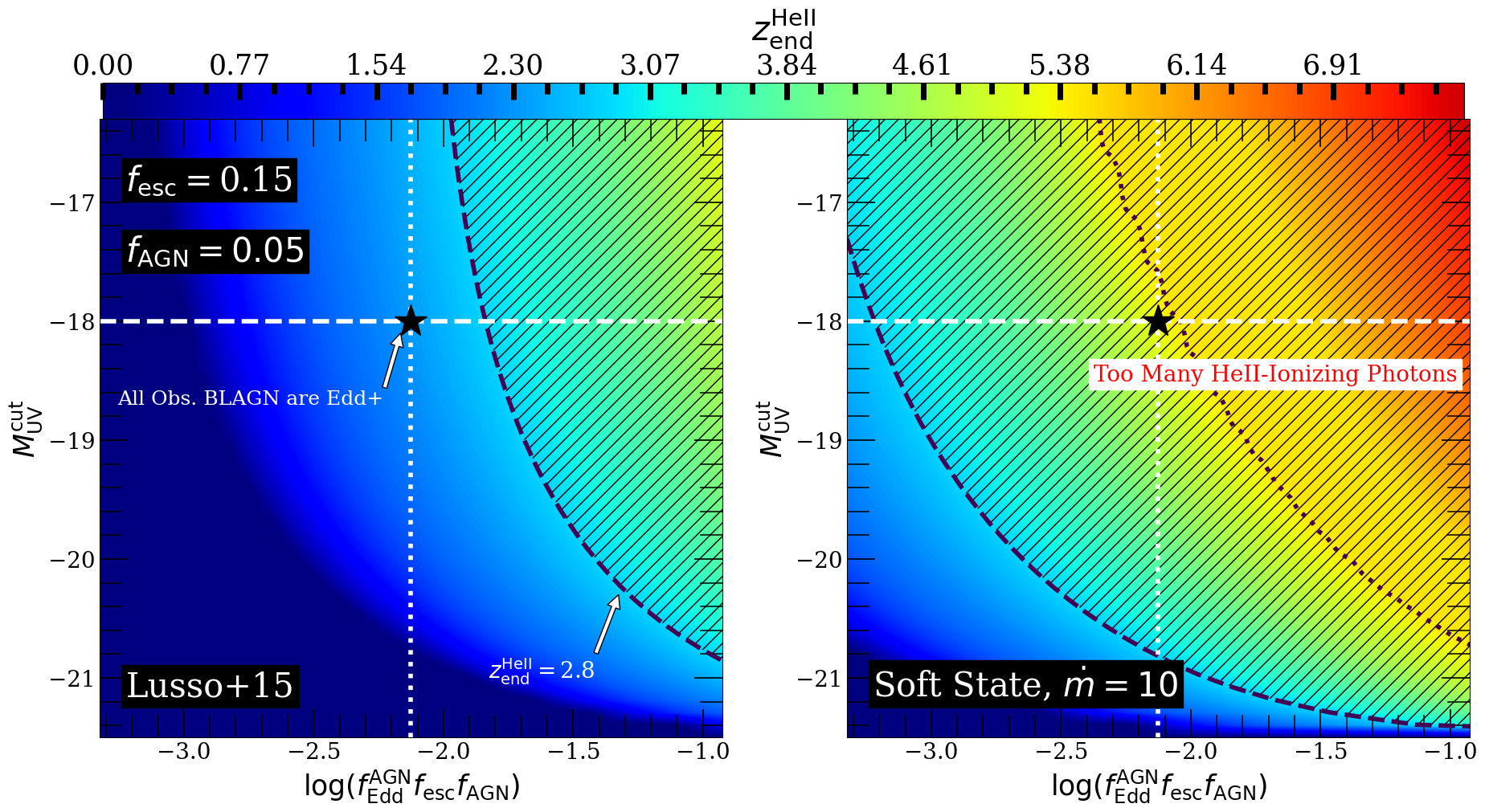}
    \caption{AGN parameter space excluded by constraints on He\,\textsc{ii} reionization.  Left and right panels show the L15 power law and soft state SED with $\dot{m} = 10$, respectively.  The white vertical dotted and horizontal dashed lines denote $(f_{\rm Edd}^{\rm AGN}, f_{\rm AGN}, f_{\rm esc}) = (1, 0.05, 0.15)$ and $M_{\rm UV}^{\rm cut} = -18$, respectively, and the black star denotes their intersection.  The color scale shows $z_{\rm end}^{\rm He\,\textsc{ii}}$, and the hatched region indicates where $z_{\rm end}^{\rm He\,\textsc{ii}} \geq 2.8$, which are disfavored by He\,\textsc{ii} reionization constraints.  The dotted black line indicates the end of H\,\textsc{i} reionization at $z_{\rm end}^{\rm H\,\textsc{i}} = 5.5$.  }
    \label{fig:constraints}
\end{figure*}

We will next consider how He\,\textsc{ii} reionization constrains the parameter space of BLAGN properties.  
For a given SED model and set of UVLF measurements, the resulting H\,\textsc{i} and He\,\textsc{ii} reionization histories can be characterized by $M_{\rm UV}^{\rm cut}$ and the degenerate combination $f_{\rm Edd}^{\rm AGN} f_{\rm esc} f_{\rm AGN}$.  
Requiring $z_{\rm end}^{\rm He\,\textsc{ii}} \leq 2.8$ then constrains this parameter space.  
We remind the reader that constraints obtained in this way are conservative, since we have assumed only super-Eddington AGN contribute significantly to the ionizing photon budget.  

The left panel of Figure~\ref{fig:constraints} shows the result of this exercise for the L15 power law.  
The black star at intersecting white dotted and dashed lines denotes the location where $M_{\rm UV}^{\rm cut} = -18$, $f_{\rm esc} = 15\%$, $f_{\rm AGN} = 5\%$, and $f_{\rm Edd}^{\rm AGN} = 1$ in our fiducial analysis.  
Note that our parameter space is defined analogously to that of Figure 2 of~\citet{Munoz2024}, but also including the factor $f_{\rm Edd}^{\rm AGN} f_{\rm AGN}$.  
The hatched region denotes regions where $z_{\rm end}^{\rm He\,\textsc{ii}} \geq 2.8$, which are disfavored by observations.  
We see that the L15 power law stays well under these limits for our fiducial parameters, permitting either a higher $f_{\rm esc}$ than we assume and/or other sources of He\,\textsc{ii}-ionizing photons.  

The picture changes significantly for the soft state super-Eddington SED (right panel), with the black star now well within the hatched region.  
Indeed, the dotted black line indicates where BLAGN are able to {\it also} complete hydrogen reionization by $z = 5.5$ on their own, and this almost intersects our fiducial scenario.  
Restoring agreement\footnote{We note that the CMB optical depth, $\tau_{\rm CMB}$, is minimally impacted provided AGN do not produce so many photons as to end hydrogen reionization too early as well.  This happens only at the extremes of our parameter space.  } with He II constraints would require brighter $M_{\rm UV}^{\rm cut}$ by about $-2.5$, a factor of $10$ reduction in $f_{\rm Edd}^{\rm AGN} f_{\rm esc} f_{\rm AGN}$, or a modestly smaller decrease in both.  
Between these possibilities, a reduction in $f_{\rm Edd}^{\rm AGN}$ and/or $f_{\rm esc}$ seems most likely, given that our fiducial choices for the other two parameters are comparatively well-motivated observationally (Figure~\ref{fig:UVLF}).  
Accommodating $f_{\rm Edd}^{\rm AGN} = 1$ for the $\dot{m} = 10$ soft state case would require $f_{\rm esc}^{\rm He\,\textsc{ii}} = f_{\rm esc}^{\rm H\,\textsc{i}} \lesssim 1.5\%$, comparable to or lower than expectations for star-forming galaxies around the same redshifts~\citep{Chisholm2022,Jaskot2024b,Jaskot2024a,Papovich2025}. 
We note that an average $f_{\rm esc}$ this small may be difficult to reconcile with $f_{\rm esc}$ measured in bright AGN~\citep{Grazian2018}. 
Equivalently, $f_{\rm Edd}^{\rm AGN} \approx 10\%$ would restore agreement for a $15\%$ escape fraction\footnote{Although note that in the limiting case in which $f_{\rm esc}$ is close to unity, $f_{\rm Edd}^{\rm AGN}$ would have to be $2-3\%$.  }.  
A more modest difference in several parameters (e.g. a factor of $\sim 2$--3 shift in all of them) would achieve the same goal.  
Overall, our findings indicate that a pure super-Eddington solution for LBD/LRD properties is disfavored by He\,\textsc{ii} reionization constraints unless escape fractions are somewhat lower than expected for faint AGN and/or current measurements of the abundance of BLAGN are over-estimates.  

\subsection{Caveats and Modeling Uncertainties}
\label{subsec:caveats}

Our findings are conditioned on several assumptions, chief among them being the AGN SED model.  
There is considerable debate in the literature regarding the physics of BH accretion near and beyond the Eddington limit, and how that physics shapes the emergent EUV spectrum.  
It has been suggested that super-Eddington AGN should exhibit virtually no intrinsic EUV emission due to the effects of photon trapping~\citep[e.g.,][]{Pognan2020}.  
Our conclusions would be much different in such a scenario, as there would be no ionizing photons available to drive He\,\textsc{ii} reionization.  
However, such a scenario would also likely be unable to explain the broad H$\alpha$ and H$\beta$ lines seen in LBDs and LRDs~\citep{Madau2026,Madau&Maiolino2026b}, which require EUV photons to escape the central BH and drive recombinations in the broad-line region.  
Among scenarios that do predict EUV emission, the exact amount that escapes the disk/torus system is sensitive to assumptions about gas physics and geometry.  
In Appendix~\ref{app:spectra}, we explore several other SED models that make different predictions for the intrinsic EUV emission of BLAGN.  
We find that while our qualitative conclusions remain the same, the required reduction in $f_{\rm Edd}^{\rm AGN}$ and/or $f_{\rm esc}$ can increase or decrease by a factor as much as $\approx 5$ depending on these assumptions.  
A more comprehensive study of the implications of different SED models is thus strongly motivated by this work.  

Another caveat is our assumption of $C = 3$ in Equation~\ref{eq:photon_counting_eqn}.  
Although a clumping factor of $\sim 3$ or not much higher has been long-assumed for the re-ionized IGM~\citep[e.g.][]{Pawlik2009,Shull2011}, several recent works have suggested somewhat higher values on various observational and theoretical grounds~\citep{Davies2024c,Austin2025}.  
Notably,~\citet{Davies2024c} reported observational constraints on $C$, and found it to be $\sim 12$, a factor of $4$ higher than the value assumed here.  
They found that increasing $C$ from $3$ to $12$ can delay H\,\textsc{i} reionization by roughly one redshift (see also~\citet{Munoz2024}), and we have checked and found that the same is true of He\,\textsc{ii} reionization.  
We found that a clumping factor of $12$ would weaken our constraints on $f_{\rm Edd}^{\rm AGN}$ by roughly a factor of $\approx 2.7$ relative to Figure~\ref{fig:constraints}.  

Lastly, as mentioned at the beginning of \S\ref{sec:constraints}, our fiducial scenario assumes that the escape fraction of He\,\textsc{ii}-ionizing photons is the same as that of H\,\textsc{i}-ionizers.  
If He\,\textsc{ii} escape fractions are significantly lower, as suggested by~\citet{Madau2024}, the tension may be weaker than we report by a factor of $\approx 2$ for a fixed H\,\textsc{i}-ionizing escape fraction.  
Indeed, in combination with a higher clumping factor and a modestly softer SED, this may be enough to alleviate the tension seen in Figure~\ref{fig:constraints}.  
It is also possible that anisotropy of the intrinsic emission of the accreting BH could result in the isotropically-averaged escape fraction being lower than one would expect based on line-of-sight measurements. 
The relationship between line-of-sight escape fraction and anisotropic intrinsic emission is strongly dependent on escape geometry, the column density of absorbing gas, and the ionizing spectrum of the source, and thus will require further investigation.  

\section{Conclusions}
\label{sec:conclusions}

We have studied the implications of constraints on the timing of He\,\textsc{ii} reionization for the abundance of super-Eddington BLAGN at $z > 3$.  
We have used realistic models for intrinsic SEDs of super-Eddington black holes, and focused on the proposed scenario in which most high-redshift BLAGN are above the Eddington limit. 
Our conclusions are: 

\begin{itemize}

    \item Our fiducial super-Eddington AGN SED model is significantly more efficient at producing He\,\textsc{ii}-ionizing photons than a standard power-law template for BH masses in the range $\sim  10^5$--$10^7~\rm{M}_{\odot}$.  
    BHs in this mass range would likely dominate the He\,\textsc{ii} ionizing emission of the BLAGN population if all of them are near or above the Eddington Limit.

    \item Constraints placing the end of He II reionization at $z\sim3$ are difficult to reconcile with a scenario where most BLAGN are super-Eddington.   
    Restoring agreement with He\,\textsc{ii} reionization constraints would require the fraction of BLAGN that are super-Eddington to be $\lesssim 10\%$ for our fiducial, observationally-motivated parameter choices.  
    Alternatively, an ionizing photon escape fraction of $\lesssim 1.5\%$ could achieve the same result, although such low escape fractions may be difficult to reconcile with observations of bright AGN.  
    
\end{itemize}

Our findings are predicated on our assumed model for the ionizing spectra of super-Eddington AGN, which predict a relatively hard EUV spectrum compared to standard power-law templates.  
Additional assumptions we made, which might weaken the tension if relaxed, include a clumping factor of $C = 3$ and equal escape fractions for H\,\textsc{i} and He\,\textsc{ii}-ionizing photons.  
Forthcoming efforts to explain the underlying physics of LRDs and LBDs should take constraints on He\,\textsc{ii} (and H\,\textsc{i}) reionization into account.  

\begin{acknowledgments}

The authors thank Tong Su for providing his code to compute AGN SEDs for the soft state model.  
We also thank Piero Madau, Roberto Maiolino and Julian Mu\~{n}oz for helpful comments and discussion that helped strengthen this manuscript.  
CC acknowledges support from the
Beus Center for Cosmic Foundations at Arizona State University. 
YZ acknowledges support from JWST/NIRCam contract to the University of Arizona NAS5-02105.  
LW acknowledges support from the Gavin Boyle Fellowship at the Kavli Institute for Cosmology, Cambridge and from the Kavli Foundation.
AD acknowledges support
from NSF grant AST-2045600. RAW acknowledges support from NASA JWST Interdisciplinary Scientist grants
NAG5-12460, NNX14AN10G and 80NSSC18K0200 from GSFC.  

\end{acknowledgments}

\bibliography{references}{}
\bibliographystyle{aasjournalv7}

\appendix

\section{Constraints for Alternative SEDs}
\label{app:spectra}

\begin{figure*}[h!]
    \centering
    \includegraphics[width=0.93\linewidth]{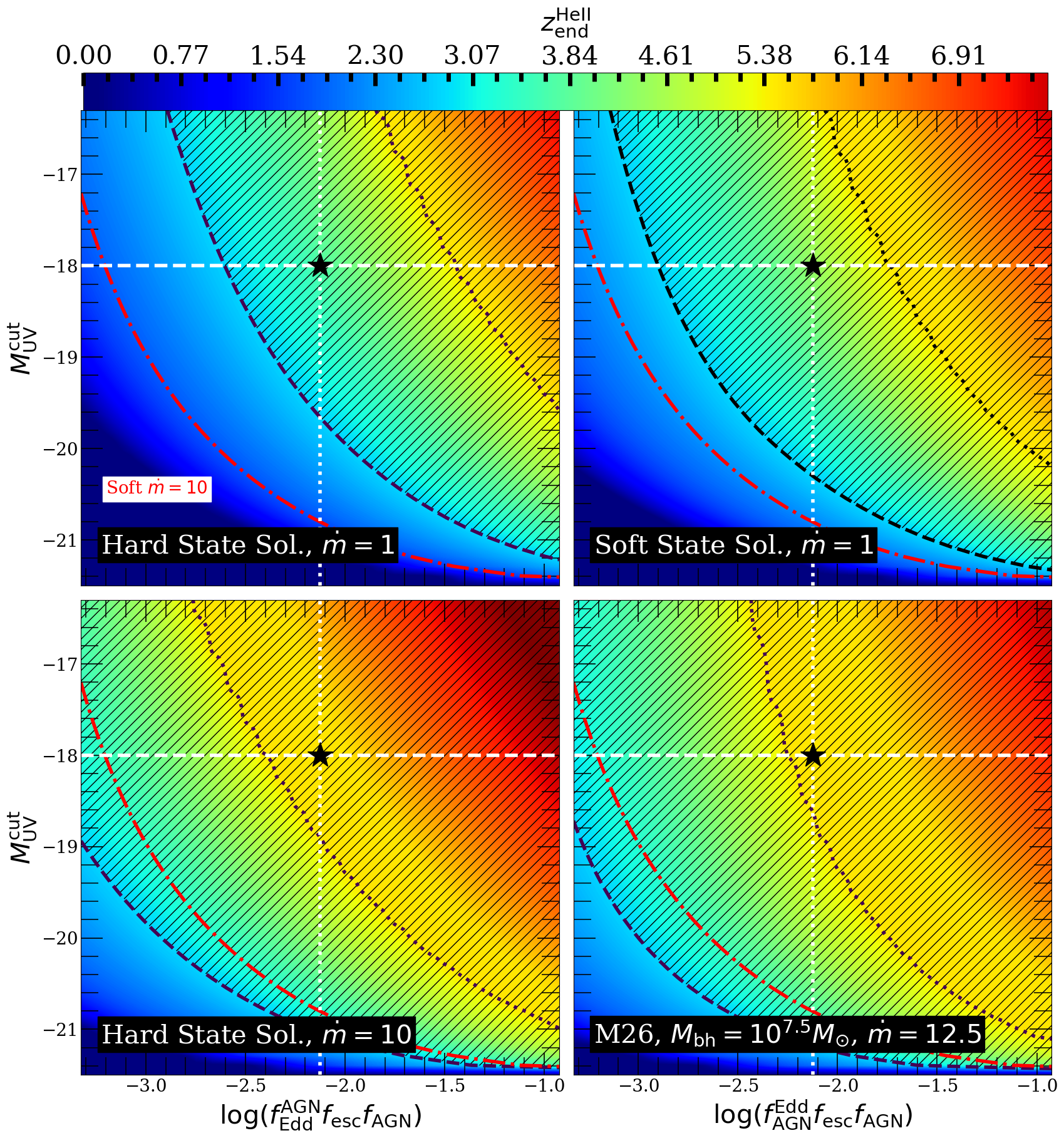}
    \caption{Constraints on AGN properties for other physically-motivated BH accretion scenarios that produce significant EUV emission.  The top and bottom left panels show the Hard State case with $\dot{m} = 1$ and $10$.  The top right shows the Soft State with $\dot{m} = 1$, and the bottom right shows the ``model B'' scenario proposed by~M26 for an $M_{\rm BH} = 10^{7.5}~\rm{M}_{\odot}$ BH with $\dot{m} = 12.5$.  In all panels, the red dot-dashed line denotes the constraint for the $\dot{m} = 10$ Soft State model shown in the right panel of Figure~\ref{fig:constraints}.  Both cases with $\dot{m} = 1$ result in weaker constraints than in the right panel of Figure~\ref{fig:constraints}, while the other two yield tighter constraints.  }
    \label{fig:alt_spectra}
\end{figure*}

We briefly consider here alternative scenarios for the SEDs of BLAGN, and how these affect our findings.  
We assume three alternative scenarios: the ``hard state'' model of~\citet{Liu2003}, our fiducial soft state spectrum with $\dot{m} = 1$, and the spectrum provided in Figure 4 of M26 for a BH with mass $10^{7.5}~\rm{M}_{\odot}$ and $\dot{m} = 12.5$ (see Figure~\ref{fig:spectra}).   
The top and bottom-left panels show results for the Hard State solution with $\dot{m} = 1$ and $\dot{m} = 10$.  
The top right shows the soft state for $\dot{m} = 1$, and the bottom right assumes that all objects have the same shape SED as the angle-average of the spectra from Figure 4 of~M26.  
We also show the dashed black line in the right panel of Figure~\ref{fig:constraints} - our fiducial constraint for the $\dot{m} = 10$ soft state model - as a red dot-dashed curve in each panel.  

Both $\dot{m} = 1$ scenarios display somewhat weaker constraints than in the right panel of Figure~\ref{fig:constraints}.  
In the Hard State case, ending He\,\textsc{ii} reionization on time would require only a factor of $2$ decrease in $f_{\rm esc}$ or $f_{\rm Edd}^{\rm AGN}$ from our fiducial value.   
In the Soft State case, this becomes a factor of $5$.  
In the bottom panels, however, constraints become stronger than in Figure~\ref{fig:constraints}.  
Both scenarios end would require over a factor of $20$ lower ionizing output, and (for our fiducial parameters) produce enough photons to complete both H and He reionization without any help from galaxies.  
We note that the M26 model may produce even more photons if the dependence of the SED on black hole mass were included self-consistently, as it is for the Hard and Soft State scenarios.  
These findings motivate a more comprehensive exploration of the parameter space of BH accretion disk models to determine which types of models (if any) can naturally accommodate a high super-Eddington AGN fraction at $z = 3$--$7$ and a late end to He\,\textsc{ii} reionization.  

\end{document}